Preparation of Thin Crystals of FeTe$_{1-x}$S$_x$ Using the Scotch-Tape Method

Yoshikazu Mizuguchi[1], Tohru Watanabe[2], Hiroyuki Okazaki[2,3], Takahide Yamaguchi[2,3], Yoshihiko Takano[2,3], and Osuke Miura[1]

[1] Department of Electrical and Electronic Engineering, Tokyo Metropolitan University, Hachioji, Tokyo, 192-0397, Japan.

[2] National Institute for Materials Science, Tsukuba, Ibaraki, 305-0047, Japan.

[3] JST-TRIP, Tsukuba, Ibaraki, 305-0047, Japan.

Abstract

We have investigated an applicability of the scotch-tape method to fabricating thin crystals of Fe chalcogenide FeTe$_{1-x}$S$_x$. Thin crystals with a typical thickness of 40 nm were expectedly left on a Si substrate. Thin films prepared using this process will develop both detailed studies on intrinsic nature of Fe-based superconductivity and application to superconducting devices.

Compounds with a layered crystal structure have attracted much attention due to the wide range of physical properties basically resulted from the two-dimensional electronic state. Superconductivity with a high transition temperature ($T_c$) often occurs in such a layered structure: copper oxides[1], boron carbides[2], HfNCl system[3], and $MgB_2$[4]. Recently-discovered Fe-based superconductors are also composed of layered structure[5-7]. The anomalous physical features, for example, magnetism, behavior under high fields, pressure effects, and paring symmetry have been discussed actively. Among the Fe-based superconductors, Fe chalcogenides, FeSe and FeTe family[8-12], are the simplest system because they are composed of a simple stacking of Fe-chalcogen layers without any interlayer elements as depicted in Fig. 1. Generally in such a layered crystal structure with a van-der-Waals gap, the single crystals can be cleaved easily. If we could obtain very thin single crystal by cleaving the single crystals as graphene[13], it will give us a greater understanding of the nature of Fe-based superconductivity and devise applications of Fe-based superconductors.

The scotch-tape method is a powerful way to obtain a thin single crystal from a layered material. As well known, graphene can be produced using this method. Recently Ye et al. reported liquid-gated interface superconductivity on an atomically flat film of ZrNCl, which was obtained using the scotch-tape method[14]. While ZrNCl is an insulator, field-effect superconductivity was observed in the thin film containing several ZrNCl layers. We have investigated an applicability of the scotch-tape method to Fe chalcogenide FeTe$_{1-x}$S$_x$, and succeeded in obtaining Fe-chalcogenide thin crystals on a Si substrate using the scotch-tape method.

Single crystals of FeTe$_{1-x}$S$_x$ were grown using the self-flux method as described in Ref. 15. The actual S concentration ($x$) of the single crystal used in this study was determined to be 0.064 by electron probe micro analyzer (EPMA). The obtained single crystal was placed on scotch tape, and then was cleaved. We repeated cleaving more than 100 times. The scotch tape with a lot of small thin crystals was pasted onto a Si substrate. Then, we pressed the outside of the tape using a spatula to enhance the bonding between the thin crystals and the substrate, which would be achieved by van-der-Waals attraction. By using this process, thin single crystals were left on the Si substrate. An optical microscope and an atomic force microscope (AFM) were used in observation and investigation of the thickness of the thin crystals.

Figures 2(a) and 2(b) show an optical microscope image and an AFM image of the thin crystals, respectively. The thickness of the cleaved crystals (A - D) was estimated to be 140 nm for A, 57 nm for B, 40 nm for C, and 48 nm for D; the thickness depended on a frequency of cleaving. The lattice constant $c$ of this crystal determined by X-ray diffraction is 0.625 nm. The thickness of 40 nm almost corresponds to the thickness of 64 layers. This strongly suggests that thin crystals with several Fe-chalcogen layers will be obtained using the scotch-tape method by modifying the fabrication process, such as changing the adhesion of scotch tape, surface of substrate, and the quality of single crystal. Furthermore, chemical compositions of the crystals would also be an important factor to realize very thin crystal. Using such a Fe-chalcogen thin film, we will be able to study a thickness dependence of superconducting properties of Fe chalcogenides. Furthermore, those thin films will develop device applications using Fe-chalcogenide superconductors.

Acknowledgements

The authors would like to thank Professor A. Kanda at the University of Tsukuba for instructing us of the scotch-tape method. This work was partly supported by Grant-in-Aid for Scientific Research (KAKENHI).

Figure captions

Fig. 1. (Color online) Schematic image of the crystal structure of FeTe. The figure was drawn using VESTA[16].

Fig. 2. (Color online) (a) Optical microscope image of the prepared FeTe$_{0.936}$S$_{0.064}$ thin crystals on Si substrate. (b) AFM image of the FeTe$_{0.936}$S$_{0.064}$ thin crystals on Si substrate.

Figure 1.

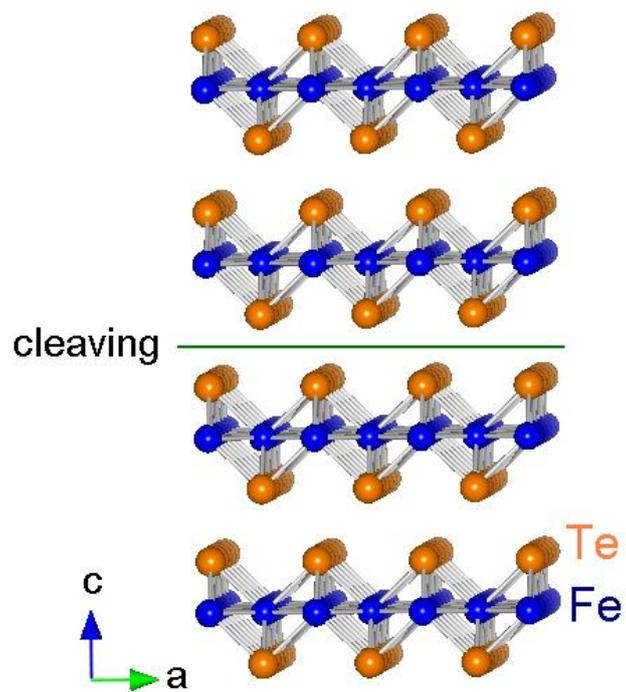

Figure 2.

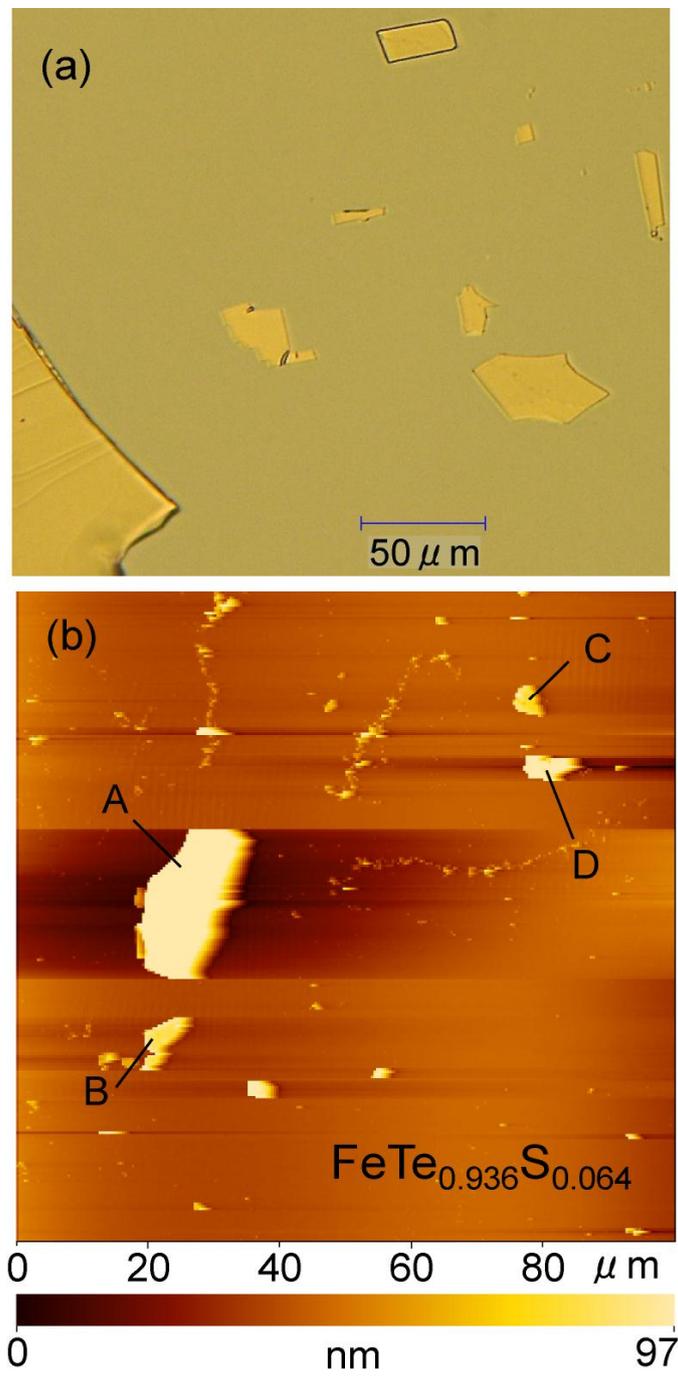